\documentclass[
showpacs,preprintnumbers,amsmath,amssymb]{revtex4}
\usepackage{amsmath}
\usepackage{empheq}
\usepackage{graphicx}
\begin{document}

\title{INFLATION IN TERMS OF A VISCOUS VAN DER WAALS COUPLED FLUID}

\author{   I. Brevik$^{1}$\footnote{E-mail:iver.h.brevik@ntnu.no}},

\author{V. V. Obukhov$^{2}$\footnote{E-mail: obukhov@tspu.edu.ru}}

 \author{ A. V. Timoshkin$^{2,3}${\footnote{E-mail:alex.timosh@rambler.ru}}}

\medskip

\affiliation{$^{1}$Department of Energy and Process Engineering, Norwegian University
of Science and Technology, N-7491 Trondheim, Norway}

\affiliation{$^{2}$Tomsk State Pedagogical University, Kievskaja Street 60, 634050 Tomsk, Russia}

\affiliation{$^{3}$National Research Tomsk State University, Lenin Avenue 36, 634050 Tomsk, Russia}

 \today

\begin{abstract}

We propose to describe the  acceleration of the universe by introducing a model  of two coupled fluids. We focus on the accelerated expansion at the early stages. The inflationary expansion  is described in terms of a van der Waals equation of state for the cosmic fluid, when account is taken of   bulk viscosity. We assume that there is a weak interaction between the van der Waals fluid and the second component (matter). The gravitational equations for the energy densities of the two components are solved for a homogeneous and isotropic Friedmann-Robertson-Walker universe, and analytic expressions for the Hubble parameter are obtained.  The slow-roll parameters, the spectral index, and the tensor-to-scalar ratio are calculated and compared with the most recent astronomical data from the Planck satellite. Given reasonable restriction on the parameters, the agreement with observations is favorable.
\end{abstract}


 \maketitle
\section{Introduction}

The experimental proof given in 1998  by Riess {\it et al.} \cite{riess98} and Perlmutter {\it et al.} \cite{perlmutter99} about  the acceleration of the present universe, has recently become supported by recent data from the Planck survey \cite{ade14}. The data point to the existence of a very early inflationary period \cite{linde08,gorbunov11}, and allow one to obtain a more detailed picture of the evolution of the early universe. The early inflationary universe is exposed to an acceleration \cite{brandenberger10} that can
 be explained within the framework of  scalar-tensor theories \cite{sahni99,bamba12}, a dark energy ideal fluid  weakly interacting with ordinary matter \cite{sahni99,bamba12,bolotin15}, or via modification of gravity \cite{nojiri17}. In the inflationary period both the total energy and the scale factor increase exponentially. The inflationary period can be attributed to perfect fluids having properties different from standard matter and radiation, by $F(R)$ gravity, or by a combination of both \cite{bamba14}. Since the stage of acceleration is the same for inflation as for the late universe, especially when the Big Rip is approached, it is possible to formulate the theory for the inflationary epoch in the same way as for  the late-time evolution. Consequently, we can for the inflationary epoch apply the perfect fluid formalism, satisfying an inhomogeneous equation of state \cite{nojiri05}.

In cosmology there are several studies of the coupling between energy and matter \cite{bolotin15}. Investigations on viscous cosmology started some time ago but the applications to the accelerating universe are rather recent. Viscous fluids can be considered as a particular case of generalized fluids \cite{capozziello06,nojiri06,nojiri07}. It is of interest to study the influence of the interaction between energy and matter on the inflation, in the presence  of a bulk viscosity. Despite the fact that the viscosity is very small in the inflationary universe,  the inclusion of viscosity allows one to get a better agreement with the Planck astronomical observations in several cases \cite{brevik16,brevik17,brevik17a}. For a recent review on the the early viscous universe, see Ref.~\cite{breviknaro17}; cf. also related papers \cite{normann16} and \cite{normann17}.

Coupled-fluid cosmological models in a hot universe, taking into account viscous properties of the fluid in a homogeneous and isotropic Friedmann-Robertson-Walker flat spacetime, were studied in Ref.~\cite{brevik16}. Inflationary cosmology for a viscous fluid in terms of the van der Waals equation of state was considered in Ref.~\cite{brevik17}. Cosmic fluids satisfying the van der Waals equation were investigated  in Refs.~\cite{jantsch16,vardiasli17}. The van der Waals model can can describe both the early and the late stages evolution of the universe. Some examples of inhomogeneous viscous fluids were considered in Refs.~\cite{elizalde14,brevik15}.

In the present paper we investigate the inflationary expansion of the early universe assuming a two-component coupled fluid model in the presence of a bulk viscosity. The first component, named the van der Waals component, is the main one,  whereas the second component (matter) is described by a homogeneous equation of state. Some variants of the form of the bulk viscosity, and the van der Waals -matter coupling, are analyzed.  The inflationary slow-roll parameters, the spectral index, and the tensor-to-scalar ratio, are considered. Restrictions on the thermodynamic parameters needed to satisfy the Planck data, are obtained. The agreement between theoretical inflationary models and the latest Planck satellite data is discussed. We assume natural units where $c=1$.

\section{Van der Waals coupled fluid for inflation in the presence of viscosity}

\subsection{Basic assumptions and analytical solutions}

In this section we will study the early-time universe, applying the formalism for an inhomogeneous viscous fluid in a spatially flat Friedmann-Robertson-Walker spacetime. As mentioned, we will describe the inflation in terms of  van der Waals equation of state parameters, and we will include a  bulk viscosity. A new element in our research is the application of the van der Waals model in the inflationary epoch when there are two coupled fluids present.

The main component in our model for the cosmic fluid, the van der Waals component, is characterized by the pressure $p$ and the energy density $\rho$. The second component is matter, with corresponding symbols $p_1$ and $\rho_1$. In our model with scale factor $a$ the background equations are \cite{nojiri05a}
\begin{align}
\dot{\rho}+3H(p+\rho)=-Q, \notag \\
\dot{\rho}_1+3H(p_1+\rho_1)=Q, \notag \\
\dot{H}=-\frac{k^2}{2}(p+\rho+p_1+\rho_1). \label{1}
\end{align}
Here $H=\dot{a}/a$ is the Hubble rate, $k^2=8\pi G$ with $G$ meaning Newton's gravitational constant, and $Q$ is a function that describes the energy exchange between the fluids. A dot denotes derivative with respect to the cosmic time $t$. The cosmological constant $\Lambda$ is set equal to zero.

The first two equations in (\ref{1}) describe the dynamics of van der Waals energy and matter, whereas the third is the acceleration equation. The system (\ref{1}) models phenomenologically the behavior of the interacting fluids, when one in addition introduces the equation of state.

We write down the flat FRW metric
\begin{equation}
ds^2=-dt^2+a^2(t)\sum_i dx_i^2, \label{2}
\end{equation}
and Friedmann's equation for the Hubble rate
\begin{equation}
H^2=\frac{k^2}{3}(\rho+\rho_1). \label{3}
\end{equation}
Let us suppose that the ratio $r=\rho_1/\rho$ is constant. This assumption is related to the coincidence problem in the $\Lambda$CDM model.  Friedmann's  equation (\ref{3}) can then be rewritten as
\begin{equation}
\rho=\frac{3H^2}{k^2(1+r)}. \label{4}
\end{equation}
We will describe the inflationary universe as a fluid obeying a nonlinear inhomogeneous equation of state,
\begin{equation}
p=\omega(\rho,t)\rho+f(\rho)-3H\zeta(H,t). \label{5}
\end{equation}
Here the thermodynamic parameter $\omega(\rho,t)$ depends on the van der Waals energy and time, the bulk viscosity $\zeta(H,t)$ depends on the Hubble parameter and time, and $f(\rho)$ is an arbitrary function in the general case. An effective equation of state of this class is typical for modified gravity \cite{nojiri03,nojiri07a}.

Let us assume that the van der Waals  fluid obeys a parameterized  equation of state, and let us choose the following  forms for the function $f(\rho)$ and  the viscosity,\begin{align}
\omega(\rho, t)=\frac{\gamma}{1-\beta \rho/\rho_c}, \notag \\
f(\rho)=-\frac{\alpha}{\rho_c}\rho^2, \notag \\
\zeta(H,t) = \tau (3H)^n. \label{6}
\end{align}
The model thus contains three independent parameters: (i) $\gamma$ and $\beta$ connected with the van der Waals equation; (ii) $\alpha$ connected with the inhomogeneity of the equation of state (\ref{5}), in turn related to the intermolecular interaction; and (iii) the exponent $n$ describing the power dependence of the viscosity on $H$.   We thus assume basically a power law for the bulk viscosity, as is quite usual in macroscopic cosmological theory. Most often, one will make the choice  $n=1$ whereby $\zeta=3\tau H$; cf., for instance, Ref.~\cite{brevik17}. Let us here, however, first make the choice $n=3$, corresponding to a heavier weight on the  $H$ dependence in the early universe, when $H$ was large. The critical value $\rho_c$ indicates the value at which a phase transition occurs. For  the van der Waals fluid, $\tau$ is a positive constant.

\bigskip
\noindent {\bf The case $n=3$}.

\bigskip
 With $\zeta (H,t)=27 \tau H^3$ a cubic function of the Hubble parameter the equation of state takes the form
\begin{equation}
p=\frac{\gamma \rho}{1-\beta \rho/\rho_c}-\frac{\alpha}{\rho_c}\rho^2-9\tau (1+r)^2k^4\rho^2. \label{7}
\end{equation}
Since the general picture of acceleration is the same, for the inflationary stage as for the late universe, especially near the Big Bang singularity, we assume the case of nonrelativistic matter  (local rest inertial frame). We assume that the matter  component is dust, so that $p_1=0$. The pressure is thus only the van der Waals fluid pressure.The gravitational equation for matter reduces to
\begin{equation}
\dot{\rho}_1+3H\rho_1=Q. \label{8}
\end{equation}

Assume now that the interaction term between fluid and matter has the form
\begin{equation}
Q=3\lambda H(\rho+\rho_1)=3\lambda (1+r)H\rho, \label{9}
\end{equation}
where $\lambda$ is a nondimensional constant. The interaction becomes thus critically dependent on the sign of $\lambda$. As there exists no fundamental theory able to describe the interaction  term in detail, one has to limit oneself to phenomenological approaches \cite{bolotin15}.

Using the first equation in (\ref{1}) and Eqs.~(\ref{4}) and (\ref{7}), one obtains the gravitational equation for the fluid energy component,
\begin{equation}
\dot{\rho}+3H\left( \frac{\gamma \rho}{1-\beta \rho/\rho_c}-\frac{\alpha}{\rho_c}\rho^2-9\tau (1+r)^2k^4\rho^2+\rho \right) = -3\lambda (1+r)H\rho. \label{10}
\end{equation}
It is  physically reasonable to assume that the intermolecular parameter $\alpha$ is related to the viscosity parameter $\tau$.  For mathematical convenience we will assume simple proportionality and write the connection as $\alpha=-9\tau \rho_c(1+r)^2k^4$. Moreover, we
introduce the symbol $\theta$ defined by $\theta=1+\lambda (1+r)$. Using the Friedmann equation (\ref{3}), we may rewrite Eq.~(\ref{10}) as
\begin{equation}
\frac{2}{3}\dot{H}+\frac{\gamma +\theta \left( 1-A H^2\right)}
{1- A H^2}H^2=0 , \quad A=\frac{3\beta}{1+r\rho_ck^2}.\label{11}
\end{equation}
The general solution of Eq.~(\ref{11}) is
\begin{equation}
(\gamma +\theta)^{-1}\left[ H^{-1}+\sqrt{\frac{A}{\theta}}(\theta-\sqrt{\gamma+\theta})\ln \frac{\sqrt{A\theta}H-\sqrt{\gamma+\theta}}{\sqrt{A\theta}H+\sqrt{\gamma+\theta}}\right] =\frac{3}{2}t+{\rm constant}. \label{11a}
\end{equation}
This equation gives the Hubble parameter as an implicit function of time.  Mainly  because of mathematical simplicity we will henceforth make the assumption  $\gamma=\theta(\theta-1)$. In that way we neglect the slowly varying logarithmic function of $H$ but keep the rapidly varying part. Physically,  this  means that the parameter $\gamma$ in the van der Waals equation (\ref{6}) is related to the interaction parameter $\lambda$  in the interaction $Q$ between the fluids; cf. Eq.~(\ref{9}).

 Then the solution of Eq.~(\ref{11a}) becomes quite simple,
\begin{equation}
H=\frac{2}{\theta^2(3t+C)}, \label{12}
\end{equation}
where $C$ is an arbitrary constant. Choosing $C=0$ we find that $H(0)$ becomes infinite at the Big Bang, $t=0$.

Solving the gravitational equation of motion (\ref{8}) for matter with the coupling term (\ref{9}) and the Hubble parameter (\ref{12}) we obtain
\begin{equation}
\rho_1(t)=\tilde{\rho}_1(3t+C)^{2(\theta-2)/\theta^2}, \label{13}
\end{equation}
where the constant $\tilde{\rho}_1$ is arbitrary.

Thus far, we have presented an example of a van der Waals viscous fluid coupled with matter, in the inflationary regime.

\subsection{Calculation of experimentally detectable parameters}

We intend now to investigate how this inflationary model conforms with recent results from the Planck satellite. Therewith appropriate constraints on model parameters can be found. From the solution (\ref{12}) we calculate the slow-roll inflationary parameter,
\begin{equation}
\varepsilon = -\frac{\dot{H}}{H^2}=\frac{3}{2}\theta^2. \label{14}
\end{equation}
In order to have a regime of acceleration we must have $\varepsilon \ll 1$, whence $\lambda (1+r) < \sqrt{2/3}-1.$ This means that the parameter $\lambda$ becomes negative. It is worth noticing that the development in the acceleration regime depends on the sign of the interaction constant.

Another important slow-roll parameter is
\begin{equation}
\eta=\varepsilon -\frac{1}{2\varepsilon H}\dot{\varepsilon}. \label{15}
\end{equation}
In our case $\eta=\varepsilon$. The power spectrum is
\begin{equation}
\Delta_R^2=\frac{k^2H^2}{8\pi^2\varepsilon}, \label{16}
\end{equation}
and with Eq.~(\ref{12}) for the Hubble parameter the power spectrum takes the form
\begin{equation}
\Delta_R^2=\frac{k^2}{3\theta^6(3t+C)^2}. \label{17}
\end{equation}
From the slow-roll parameters we can calculate the spectral index $n_s$ and the tensor-to-scalar ratio $r$,
\begin{equation}
n_s=1-6\varepsilon +2\eta, \quad r=16\varepsilon, \label{18}
\end{equation}
and obtain
\begin{equation}
n_s=1-6\theta^2, \quad r=24 \theta^2. \label{19}
\end{equation}
From the Planck data it is known that $n_s=0.9603 \pm 0.0073.$ In order to comply with this result, we must require that $\theta \approx 0.08134$. From the Planck data one has further that $r<0.11$. In our model $r = 24\theta^2 \approx 0.159$, so that the inequality is actually violated.

On the other hand, if we take into account the recent BICEP2 data \cite{ade14a} according  to which $r=0.20^{+0.07}_{-0.05}$, we see that the present model can comply with inflation.

Notice that if we were to compare the theoretical inflation tensor-to-scalar value for a one-component fluid, with coupling constant $\lambda=0$, then we would get  $\theta=1$. In this case the inflationary model would be unsuitable.

\bigskip
{\bf The case $n=1$}.

\bigskip

Let us consider another one-parameter van der Waals equation of state, making the choices   $\alpha=9\gamma/8, \beta=1/3$  (not related to the choice for $\alpha$ made after Eq.~(\ref{10})), and let us choose the more traditional   value $n=1$ in Eq.~(\ref{6}) for the bulk viscosity. Then  $\zeta(H,t)=3\tau H$ becomes a linear function of the Hubble parameter. As before, we take the coupling to be as in Eq.~(\ref{9}). The van der Waals equation takes the form
\begin{equation}
p=\frac{\gamma \rho}{1-\frac{1}{3}\frac{\rho}{\rho_c}}-\frac{9}{8}\frac{\gamma}{\rho_c}\rho^2-3\tau (1+r)k^2\rho. \label{20}
\end{equation}
Taking into account Eqs.~(\ref{1}), (\ref{4}) and (\ref{20}) we arrive at the following differential equation for the energy density,
\begin{equation}
a\left( 1-\frac{1}{3}\frac{\rho}{\rho_c}\right) \frac{d\rho}{da}+3\rho \left[\gamma+\tilde{\theta}-\left( \frac{1}{3}\tilde{\theta}+\frac{9}{8}\gamma \right) \frac{\rho}{\rho_c}+\frac{3}{8}\left( \frac{\rho}{\rho_c}\right)^2 \right] =0, \label{21}
\end{equation}
where $a$ is the scale factor and $
\tilde{\theta}=(1+r)(\lambda-3\tau k^2)+1$.

Using the Friedmann equation (\ref{4}) and introducing the scale variable $x=\rho/\rho_c$ we can express Eq.~(\ref{21}) in the form
\begin{equation}
\frac{3 da}{a}=- \frac{\left( 1-\frac{1}{3}x\right)dx}{x(Ax^2+Bx+C)}. \label{22}
\end{equation}
We have here introduced $A=\frac{1}{8}\gamma, B=-\left( \frac{3}{8}\gamma+\frac{1}{9}\tilde{\theta}\right), C=\gamma+\tilde{\theta}$.

The solution of Eq.~(\ref{22}) depends on the sign of the discriminant  $4AC-B^2$ of the polynomial $Ax^2+Bx+C$. For simplicity we will only consider the case where the discriminant is equal to zero. Then the solution of Eq.~(\ref{22}) becomes
\begin{equation}
a=a_0\left[\frac{x+\frac{B}{2A}}{x}\right]^{\frac{4A}{3B^2}}\exp \left[ -\frac{\frac{1}{3A}+\frac{2}{B}}{3\left( x+\frac{B}{2A}\right)}\right]. \label{23}
\end{equation}
In order to simplify this expression we make the choice $B=-6A$. Then $\gamma= 8\tilde{\theta}/27$, and we obtain
\begin{equation}
a=a_0\left( \frac{|x-3|}{x}\right)^{1/\tilde{\theta}}. \label{24}
\end{equation}
In the beginning of inflation, when $a\rightarrow 0$, the initial value of the density is $\rho_{\rm in}=3\rho_c$.

We next compare how this theoretical model compares with the latest Planck data. We use the scale factor (\ref{24}) and calculate the slow-roll parameters $\varepsilon$ and $\eta$, \begin{equation}
\varepsilon =\frac{1}{2}\tilde{\theta}\frac{|x-3|}{|x-3|-x}, \quad \eta=\frac{1}{2}\tilde{\theta}. \label{25}
\end{equation}
The initial values of the parameters are $\varepsilon (3\rho_c)=0$ and $|\eta(3\rho_c)|=\frac{1}{2}\tilde{\theta}$. The first  condition $\varepsilon \ll 1$ for inflation is thus fulfilled. The second condition $\eta \ll 1$ leads to the inequality $\tilde{\theta} \ll 1$. One obtains the result $(1+r)(\lambda-3\tau k^2) \ll 1$. Since the interaction between fluid energy and matter is assumed to be weak, the constant $\lambda$ is very small, and the second term in the parenthesis  is also small since it contains $k^2$.  The requirement on  the slow-roll parameter $\eta$ can thus be satisfied.

The spectral index is given by
\begin{equation}
n_s-1=-\tilde{\theta}\left( 2+\frac{x}{|x-3|-x}\right). \label{26}
\end{equation}
With the condition $r<0.11$ we have $\frac{x}{|x-3|-x}<0.01375\,{\tilde{\theta}}^{-1}-1$ and $n_s-1 \approx -0.0(3)<-\tilde{\theta}(1+0.01375 \,{\tilde{\theta}}^{-1})$, or equivalently $\tilde{\theta}<0.31958$. Thus, in order to correspond with the Planck results it is necessary to require $\gamma < 0.0947$. Note that by assuming instead a nonviscous one-component fluid, with $\tau=0$ and $\lambda=0$, we would obtain $\gamma \approx 0.296$. The Planck results would not be in agreement with such a model.

We may now solve the gravitational equation of motion (\ref{8}) for matter, with the coupling (\ref{9}) rewritten in the form
\begin{equation}
Q=3\lambda \left( 1+\frac{1}{r}\right) H\rho_1. \label{27}
\end{equation}
Taking into account the law of evolution (\ref{24}) for the scale factor we can calculate the Hubble parameter
\begin{equation}
H=\frac{\dot{a}}{a}={\tilde{\theta}}^{-1}\, \frac{x-|x-3|}{x|x-3|}\dot{x}. \label{28}
\end{equation}
From the Friedmann equation (\ref{3}) we obtain the matter density,
\begin{equation}
\rho_1=\frac{3r}{1+r}\frac{1}{(k\tilde{\theta})^2}\left( \frac{x-|x-3|}{x|x-3|}\dot{x}\right)^2. \label{29}
\end{equation}
Then the gravitational equation for matter takes the form
\begin{equation}
\ddot{x}-\delta \frac{|x-3|-x}{x|x-3|}{\dot{x}}^2=0, \label{30}
\end{equation}
with the notation $\delta=1+\frac{3}{\tilde{\theta}}\left( 1-\lambda \frac{r+1}{r}\right).$

We find the solution
\begin{equation}\delta \ln \frac{3}{x}+\frac{x}{3}-\sum_{k=1}^{\delta-1}\frac{k}{\delta-k}\left( \frac{x-3}{x}\right)^{\delta-k}=C_1t+C_2, \label{31}
\end{equation}
where $C_1$ and $C_2$ are arbitrary constants.Here it is taken into account that that the inflation takes place when the energy density is $\rho >\rho_{\rm in}=3\rho_c~(x>3)$.

Now rewrite Eq.~(\ref{31}) in terms of the matter density,
\begin{equation}
\frac{\rho_1}{3r\rho_c}+\delta \ln \frac{3r\rho_c}{\rho_1}-\sum_{k=1}^{\delta-1}\frac{k}{\delta-k}
\left( 1-\frac{3r\rho_c}{\rho_1}\right)^{\delta-k}=C_1t+C_2. \label{32}
\end{equation}
In particular, if $\lambda=r/(r+1)$ we obtain
\begin{equation}
\frac{\rho_1}{3r\rho_c}+\delta \ln \frac{3r\rho_c}{\rho_1}=C_1t+C_2. \label{33}
\end{equation}
Thus, we have shown how inflation can be realized by making use of the van der Waals viscous coupled fluid model.

\section{Conclusion}

 In this paper  we investigated a   van der Waals  cosmological model for inflation  in the presence of viscosity, in a flat  Friedmann-Robertson-Walker spacetime. We assumed an interaction $Q$ (cf. Eq.~(\ref{9})) to take place between two coupled fluids, one viscous van der Waals component, and one pressure-free matter  component, and studied  the influence from this interaction  on the inflationary parameters. In this way we found solutions of the gravitational equations of motion in the parameterized   van der Waals model. Three constant independent parameters, called $\alpha, \beta$ and $\gamma$ (cf. Eq.~(\ref{6})) were contained  in the model. From these fundamental parameters the   parameters of experimental interest were calculated, such as  the spectral index $n_s$ and the tensor-to-scalar ratio $r$. The viability of the model was tested via comparison between theoretical inflationary parameters and the latest data observed from the Planck satellite. The analysis showed that by some restrictions on the parameters in the van der Waals equation of state we obtained good agreement with the observations.

 \bigskip

 Because of mathematical tractability we made some simplifying assumptions in the formalism. It may be convenient to collect  the essential ones again here, and at the same time  mention their physical relationships:

 \noindent 1. Motivated by the coincidence problem in $\Lambda$CDM cosmology, we took the ratio between the fluid densities $r=\rho_1/\rho$ to be constant.

 \noindent 2. The bulk viscosity $\zeta(H,t)$ was assumed to follow a power law; cf. Eq.~(\ref{6}). In the first part of the paper we chose $n=3$ for the exponent, meaning physically that the influence from bulk viscosity was taken to be  large at the beginning of the universe's evolution. In the second half of the paper, we assumed a milder and more conventional form, $n=1$.

 \noindent 3. In connection with Eq.~(\ref{10}), we took the intermolecular parameter $\alpha$ to be related to the basic viscosity parameter $\tau$; cf. Eqs.~(\ref{6}). From a physical viewpoint this is reasonable. As a simplest choice we assumed a direct proportionality, $\alpha \propto \tau$.

 \noindent 4. When solving the  complicated equation (\ref{11a}) for the Hubble parameter with respect to time, we made the choice $\gamma=\theta(\theta-1)$ for the van der Waals parameter $\gamma$, whereby we kept the rapidly varying part in $H$ but omitted the slowly varying logarithmic part. Physically this meant that $\gamma$ was taken to be connected with the interaction parameter $\lambda$ in the exchange term $Q$; cf. Eq.~(\ref{9}).

 \bigskip

 An analysis of inflation in the presence of two coupled fluids, assuming a flat Friedmann-Robertson-Walker universe, was considered earlier in Ref.~\cite{brevik16}. As was shown there, the addition of a hypothetical second fluid in the theoretical inflationary models allows one to improve the agreement with astronomical observations. The inflationary expansion of the early-time universe in terms of the van der Waals equation of state in the presence of a bulk viscosity was considered in Ref.~\cite{brevik17}, with the conclusion that the inclusion of viscosity in the inflationary epochs tends to improve the cosmological models. The present study is motivated by our desire to combine the theoretical ideas of these two works. Our main result is that the possibility to describe the inflationary universe in terms of two-components fluids is a viable one.



\end{document}